# Orienting oxygen vacancies for fast catalytic reaction

Hyoungjeen Jeen†, Zhonghe Bi†, Woo Seok Choi, Matthew F. Chisholm, Craig A. Bridges, M. Parans Paranthaman, and Ho Nyung Lee*

Oak Ridge National Laboratory, Oak Ridge, TN, 37831, USA
[†] These authors contributed equally to this work.
[*] E-mail: hnlee@ornl.gov

Oxygen reduction reaction (ORR) is one of the most critical rate-limiting processes in many energy and environmental technologies, including catalytic converters, oxygen-separation membranes, solid oxide fuel cells (SOFC), and batteries[1-5]. In order to facilitate the catalytic reaction, previous studies have focused on creating catalytically active surfaces. In the case of oxide-based catalysts, examples include the modification of surface states by creating oxygen vacancies via heavy doping of cations[6], surface segregation of cationic dopants[7, 8], introducing heterophase boundaries[9], and layer ordering[10]. However, due to the lack of single crystalline materials, the complex functional coupling between the ORR and the surface state, which can be modified by changes in, e.g. the local crystallographic symmetry, strain, and orientation, has not been well understood. The lack of such systematic studies has consequently hindered the development of new materials and strategies for advanced energy systems and technologies.

Together with cation doped mixed-valence oxides, structurally layered materials, such as oxygen vacancy ordered brownmillerites ($ABO_{2.5}$), double perovskites ($A_2BB'O_6$), and Ruddlesden-Popper compounds ($A_{n+1}B_nO_{3n+1}$), are promising candidates for catalysts due to the multivalent nature of B-site cations and useful oxygen frameworks from the anisotropic crystal structures. Among them, the brownmillerite structure features one-dimensional oxygen vacancy channels (OVCs), stemming from the alternate stack of octahedral and tetrahedral sub-layers[11-14]. Note that the OVCs run parallel to the [010] direction (see Supporting Information for crystallographic details). Owing to the intriguing structure, when a multivalent cation, e.g. Co, Ni, or Fe, is present in the layers, efficient charge transfer for the oxygen reduction can be facilitated. Moreover, a reduction in the overall lattice oxygen concentration yields the presence of OVCs and the change in crystallographic symmetry as compared to the perovskite ($ABO_3$) parent structure. This anisotropic crystal structure can provide a catalytically useful open-structured surface, when the orientation of the OVCs is controlled so as to optimize catalytic activity and ionic transport. Brownmillerite $SrCoO_{2.5}$ (BM-SCO) represents a model material to study the orientation dependent surface catalytic reaction. In particular, it is noteworthy that the formation of the brownmillerite phase ($SrCoO_{2.5}$) is thermodynamically far more favorable than that of the perovskite phase ($SrCoO_3$) under typical synthesis conditions, highlighting the robustness of the ordered vacancy channels in this system[12, 15]. Furthermore, the selective exposure of oxygen vacancies by orientation control can provide direct evidence on the role of oxygen vacancies and surface structure in the surface oxygen exchange process[6]. In this work, we unveil a dramatically enhanced catalytic activity in BM-SCO films through the epitaxial orientation control of OVCs and the surface oxygen vacancy concentration. The highly ordered OVC framework is found to provide a viable means to control the ORR.

We grew 50-nm-thick BM-SCO epitaxial thin films on (001)- and (111)-oriented yttria stabilized $ZrO_2$ (YSZ) substrates (surface area = 5 × 5 $mm^2$ and thickness = 0.5 mm) by pulsed laser epitaxy[13, 14, 16]. See **Figure 1**a, b for schematics. Prior to the BM-SCO growth, 4 nm-thick Gd-doped $CeO_2$ (GDC) buffer layers were grown on YSZ substrates. The GDC layers were used to inhibit the chemical inter-diffusion between the YSZ electrolyte and the BM-SCO cathode, without disturbing the epitaxial growth. In order to confirm the epitaxy of the thin films, we examined the samples by x-ray diffraction (XRD) $\theta$-$2\theta$ and $\phi$ scans with a four-circle x-ray diffractometer. As shown in Figure S1, (001)- and (114)-oriented single crystalline BM-SCO films were successfully grown on (001)- and (111)-oriented YSZ substrates, respectively (see Supporting Information for more details on the epitaxial relationship). These two orientations provide an ideal platform to compare the orientation-dependent ORR activity, as the (001)-oriented film has the OVCs running in-plane ($OVC_{IP}$), whereas the (114)-oriented film features the OVCs running out-of-plane ($OVC_{OOP}$), i.e. tilted by ~$60^o$ away from the substrate normal. As a consequence, the former film has no OVCs open to the surface. The latter film, on the other hand, has OVCs terminated at the surface that provide a high density of surface oxygen vacancies (see **Figure 1**). A Z-contrast scanning transmission electron microscopy (Z-STEM) image taken from a (114)-oriented BM-SCO film clearly represents the orientation of the $OVC_{OOP}$ together with the sharp interface between the GDC buffer layer and the YSZ substrate as shown in **Figure 1**c. It is also worthy to note that the surface roughness of the films was less than 1 nm, regardless of the substrates.

In order to check the OVC termination effect on the ORR, we employed electrochemical impedance spectroscopy (EIS) measurements. This approach can separate time-scaled charge/mass transport[17]. The impedance data were collected along the direction normal to the film surface in order to investigate the role of surface oxygen vacancies supplied from the OVCs. We used the three-electrode method schematically shown in Figure S2a to exclude the contribution of the Pt counter electrode (see Figure S2c for details). Thus, our EIS data include impedance contributions from the YSZ electrolyte, electrode/electrolyte interface, and ORR at the BM-SCO electrode as schematically shown in Figure S2b for our equivalent circuit model. In order to specifically extract the ORR component from impedance spectra[18], we conducted isothermal oxygen partial pressure ($P(O_2)$) dependent impedance measurements on the BM-SCO thin films with two different OVC orientations at 550 $^{o}$C, as shown in **Figure 2**a, b. We note that the annealing process for electrochemical measurements did not significantly alter the surface morphology. Upon close inspection of the Nyquist plots, we found three features that contribute to the spectra: (1) The real part of the impedance starts from a non-zero value at high frequency over the entire $P(O_2)$ range employed ($10^{-5} \leq P(O_2) \leq 1$ atm) (see Figure S2d). This value was constant regardless of the $P(O_2)$ conditions used, but changed for temperature-dependent measurements as shown in Figure S2e. Hence, we define this non-zero resistance at the high frequency regime as $R_{YSZ}$, which mainly originates from the YSZ electrolyte. (2) Although it is not significantly pronounced, the mid-frequency characteristics (i.e., the non-linear parts in Figure S2d, e) originate from the electrode-electrolyte resistance ($R_{INT}$). (3) Most importantly, the low frequency regime can be attributed to the ORR at the surface of the BM-SCO electrode, since it progressively increases as $P(O_2)$ decreases, typical for the ORR activity in cathodes, as shown in Figure 2a, b .

Based on our assignment above, we compared the $P(O_2)$-dependent ORR activity between the two BM-SCO films with $OVC_{IP}$ and $OVC_{OOP}$, in order to check the viability of OVC orientation for improved ORR. As shown in Figure 2, upon decreasing $P(O_2)$, we observed progressive changes in the area-specific resistance (ASR) and peak frequency ($f_{peak}$). Note that the ASR of the ORR part is defined as the difference in real impedance values at the mid- and low-frequency ends of the dominant arcs. The ASR progressively increased in both $OVC_{IP}$ and $OVC_{OOP}$ samples as the $P(O_2)$ decreases (see Figure 2c). Interestingly, the ASR value of the $OVC_{IP}$ BM-SCO film was more than an order of magnitude larger than that of the $OVC_{OOP}$ sample especially at high $P(O_2)$, indicating that the ORR is much more active in the $OVC_{OOP}$ sample. In addition, by plotting the imaginary component of the impedance data as a function of frequency ($-Z_{Im}$ *vs*. $f$), we could obtain the peak frequency, which is the inverse of the mean relaxation time. Here, a higher peak frequency ($f_{peak}$ = 4.9 Hz) was observed from the $OVC_{OOP}$ sample as compared to that from the $OVC_{IP}$ sample ($f_{peak}$= 0.16 Hz) at 1 atm as shown in Figure 2c. The enhanced peak frequency and the reduced ASR in the $OVC_{OOP}$ sample are consistent in confirming that the orientation of OVC plays a critical role for the ORR. In addition, as shown in Figure 2a, b, decreasing $P(O_2)$ (i.e., more reducing conditions) resulted in an increase in the ASR and a decrease in the peak frequency for both samples. This trend indicates that the reduction process was highly retarded at low $P(O_2)$ by a reduced probability of oxygen incorporation on the surface required for ORR, especially for the $OVC_{IP}$ sample.

Given that ORR activity is enhanced in a more oxidizing environment, we performed temperature-dependent impedance measurements at relatively high oxygen pressure (i.e., in air) to elucidate the temperature-dependent catalytic behavior. **Figures 3**a, b show impedance spectra from the $OVC_{IP}$ and $OVC_{OOP}$ BM-SCO films, respectively, recorded at various temperatures ($500 \leq T \leq 600$ °C). Significantly smaller ASR values were observed from the $OVC_{OOP}$ film than those from the $OVC_{IP}$ film, as summarized in Figure 3c. It is interesting to note that the ASR values of the $OVC_{OOP}$ BM-SCO film are an order of magnitude smaller than those of the $OVC_{IP}$

BM-SCO film for the entire temperature range. Similarly, we also observed greatly increased peak frequencies from the $OVC_{OOP}$ as shown in Figure 3c. For example, $f_{peak}$ of $OVC_{OOP}$ is 3.98 Hz at 550 °C, whereas $f$ = 0.04 Hz for the $OVC_{IP}$ sample, again implying the greatly improved ORR activity in the $OVC_{OOP}$ film. The frequency is at least 60 times higher than that from the $OVC_{IP}$ BM-SCO film across the entire temperature range.

The surface oxygen exchange coefficient ($k$) is closely related with the ASR and peak frequency and is an indicator of the active oxygen exchange flux on the cathode surface[3]. **Figure 4** summarizes the orientation dependent surface exchange coefficient in epitaxial BM-SCO thin films. We estimated electrical ($k^q$) and chemical ($k_{chem}$) surface exchange coefficients as well as the thermal activation energy ($E_a$) from the impedance spectra using the following Eqs. (1) – (3)[19, 20]:

$$k_{chem} = \frac{l}{\tau} \tag{1}$$

$$k^q = \frac{RT}{4F^2 A_{elec} R_{ORR} c_0} \tag{2}$$

$$R_{ORR} = \exp\left(-\frac{E_a}{k_B T}\right) \tag{3}$$

where $l$ is the thickness of BM-SCO films, $\tau$ is the inverse of the peak frequency from the Nyquist plot, $R$ is the universal gas constant, $T$ is the temperature, $F$ is the Faraday's constant, $R_{ORR}$ is the real part of the impedance for the ORR, $A_{elec}$ is the electrode area, and $c_0$ is the lattice oxygen concentration in BM-SCO calculated by taking into account the room temperature molar volume and lattice oxygen. Note that heating the samples for the EIS measurements may alter their oxygen contents. Thus, we have also examined the impact of varying the $c_0$ value up to the fully oxidized perovskite phase in order to take into account the possible oxidation of the brownmillerite films during electrochemical measurements[13, 14]. This oxygen content variation resulted in a negligible change in $k^{q}$, suggesting that the oxygen exchange activity is not significantly influenced by oxygen concentration variation within the films during the EIS measurements.

As shown in Figure 4a, $k$ of the $OVC_{OOP}$ sample is larger than that of the $OVC_{IP}$ sample across the entire $P(O_2)$ range studied. This trend implies that the surface of the sample with $OVC_{OOP}$ is catalytically more active than that of the $OVC_{IP}$ sample. Furthermore, we compared the $k$ values as a function of temperature to check the feasibility of low temperature SOFCs with BM-SCO epitaxial films. As one can find in Figure 4b, the $k_{chem}$ value of the $OVC_{OOP}$ sample is at least 60 times larger than that of the $OVC_{IP}$ sample. Given that $k_{chem}$ represents the surface oxygen exchange rate, it is possible to relate this quantity to catalytic activity. The greater catalytic activity in the $OVC_{OOP}$ sample can be attributed to the reduced energy barrier, as we have also observed a 25% reduction in the thermal activation energy ($E_a$). The ORR activation energy from the $OVC_{OOP}$ sample was $E_a$ = 1.2 eV, whereas it was $E_a$ = 1.5 eV for the $OVC_{IP}$ sample. Note that the differences in the surface exchange rate and thermal activation energy indicate that optimizing the orientation of an epitaxial film is a promising approach for SOFC applications, especially at low temperatures. Such improvements have been mainly observed through chemical doping[6, 21], often accompanies concerns about the homogeneity of a materials' functionality.

We have clearly observed a possibility of improved catalytic activity at the BM-SCO thin film surface by controlling the crystallographic orientation of OVCs in the brownmillerite lattice. As our approach involves no differences in chemical composition for a given measurement condition, we can attribute the highly anisotropic ORR activity directly to the characteristic layered structure of the brownmillerite phase. It is also worthwhile to compare our result with that of conventional perovskites, e.g. $La_{1-x}Sr_xCoO_{3-\delta}$, which are mixed electronic and ionic conductors. In these perovskites, when oxygen vacancies are created, they are randomly distributed. Therefore, the crystallographic orientation effect of cubic $La_{0.8}Sr_{0.2}CoO_3$ on the ORR was found to be insignificant in such a case – the impedance difference was less than four times[22, 23], whereas we observed an approximately 100-fold difference for the brownmillerite phase. Microscopically, the Co ions in tetrahedral and octahedral sites have different valence states, i.e. $Co^{2+}$ and $Co^{4+}$, respectively. Thus, the brownmillerite structure can accommodate additional oxygen in their tetrahedral lattice sites in conjunction with oxidation of the $Co^{2+}$ ions[13, 24, 25]. Therefore, it can be suggested that the Co ions in the tetrahedral sites have a sufficiently small thermodynamic barrier to provide electrons to nearby atomic oxygen. On the other hand, the Co ions in the octahedral site cannot accommodate more oxygen due to the absence of vacancy sites, and the $Co^{4+}$ cations will not readily oxidize further; therefore, the charge transfer process for reducing atomic oxygen should be hindered. We thus conclude that the exposure of $CoO_4$ tetrahedra at the surface via orientation control is responsible for the enhanced catalytic activity. It is worth mentioning that the open framework of the tetrahedral sublayers may also greatly help facilitate the bulk diffusion in the $OVC_{OOP}$ sample. However, due to the limited film thickness, it is impossible to extract the oxygen diffusion component from our electrochemical impedance spectra. Therefore, although we did not consider the oxygen diffusion as a primary factor modifying ORR activation energy, further investigation with bulk crystals may provide insight into the contribution of oxygen ion mobility in the channels.

In summary, epitaxial orientation control of OVCs in brownmillerite SCO thin films on ionically conducting YSZ substrates was successfully demonstrated by pulsed laser epitaxy. We observed a dramatically enhanced catalytic reaction from the (114) oriented epitaxial film, which has the OVCs terminated at the surface along with the presence of tetrahedrally coordinated $Co^{2+}$. Therefore, we believe the orientation control of the OVC for developing novel oxide based catalysts can be a viable strategy. Ultimately, our observation clearly demonstrated the importance of the epitaxial orientation control to discover new pathways to advanced energy materials with greatly improved catalytic performances.

EXPERIMENTAL SECTION.

*Thin film growth*: $SrCoO_{2.5}$ (50 nm)/Gd:$CeO_2$ (4 nm) films were grown consecutively on both (001) and (111) YSZ substrates by pulsed laser epitaxy (PLE). The growth temperature, oxygen partial pressure, laser fluence, and repetition rate were fixed at 750 °C, 100 mTorr, 1.7 J $cm^{-2}$, and 5 Hz, respectively. The sample structure and crystallinity were characterized by high-resolution four circle x-ray diffraction (XRD) (X'Pert, Panalytical Inc.). The Z-contrast STEM image was collected using Nion UltraSTEM200 operated at 200 keV.

*Electrochemical impedance spectroscopy*: A three-electrode system, including BM-SCO work electrode, symmetrical porous Pt counter electrode, and reference electrode, was employed for electrochemical impedance spectroscopy measurements. Chemical etching with a KI+HCl solution was performed to selectively remove BM-SCO and GDC as well as to make space for reference electrodes. Pt mesh with Pt wires was pasted onto the electrode surface to complete the electrical connections. The assembled cell was placed in a quartz reactor, which was supported in a tubular three-zone furnace. The assembled cell was then heated to 600 °C and held at this temperature for 24 hours in air to remove the organic solvents. The $P(O_2)$ was controlled by changing the ratio of pure Ar and $O_2$ flows, with a total flow rate of 50 ml/min. The $P(O_2)$ value was monitored with an Oxygen Sensor System (Imtech Co., USA). Impedance spectra were recorded at temperature intervals down to 500 °C in air. Subsequently, the $P(O_2)$ dependent impedance spectra were recorded at 550 °C. The measurement was performed in the frequency range 0.005 ~ $10^6$ Hz with a signal amplitude of 10 mV using VersaSTAT 4 (Princeton Applied Research), which has an internal frequency response analyzer. In order to ensure reproducibility and prevent potential misinterpretation of data related to unstable temperature and $P(O_2)$ conditions, we repeated several measurements after achieving stable $P(O_2)$ and temperature. We were able to reproduce similar spectra for over multiple cycles, which indicate that the degradation of the samples was minimal. *Zview* software was used to fit the acquired impedance data to an equivalent circuit, as shown in the inset of Figure S2b. Note that we used a constant phase element in our circuit model due to depressed semicircles from our Nyquist plots, which might originate from a distribution of time constants in individual reaction processes.

ACKNOWLEDGEMENTS

The work was supported by the U.S. Department of Energy, Basic Energy Sciences, Materials Sciences and Engineering Division.

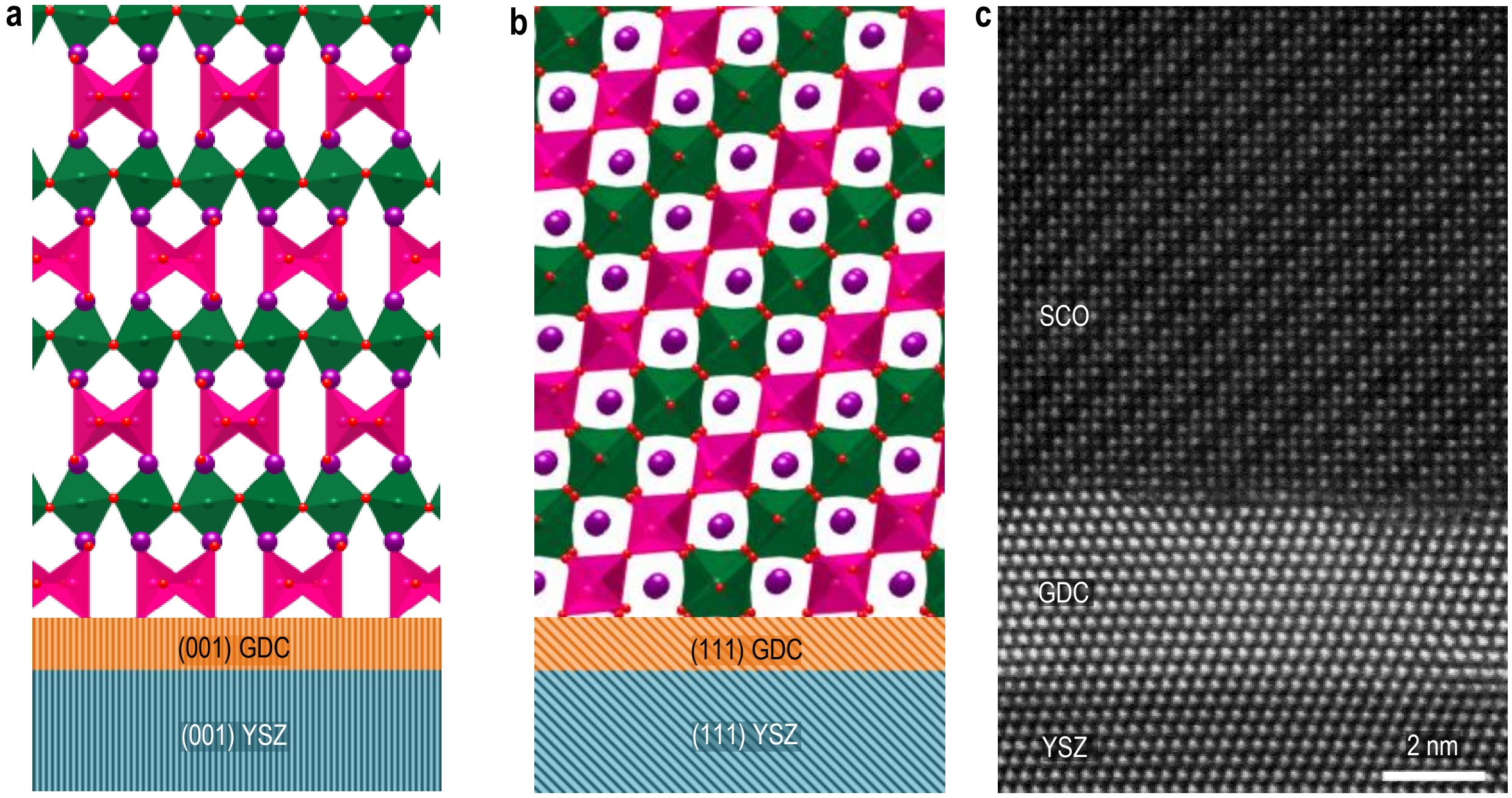

**Figure 1.** Orientated brownmillerite $SrCoO_{2.5}$ epitaxial films. Schematics of a) (001)- and b) (114)-oriented BM-SCO epitaxial films on GDC-buffered (001) and (111) YSZ substrates, respectively. The OVCs run along the [010] direction. Therefore, the (001) BM-SCO film has the OVC running along the in-plane direction, and the latter film along the out-of-plane direction. Note that for the latter case the tetrahedral sub-layers and OVCs are tilted by 45 and 60$^{o}$ away from the substrate normal, respectively. c) Cross-sectional Z-contrast STEM image of a (114)-oriented BM-SCO film on a GDC-buffered (111) YSZ substrate seen along the $[\bar{1}10]$ YSZ direction, clearly confirming the orientation of the film.

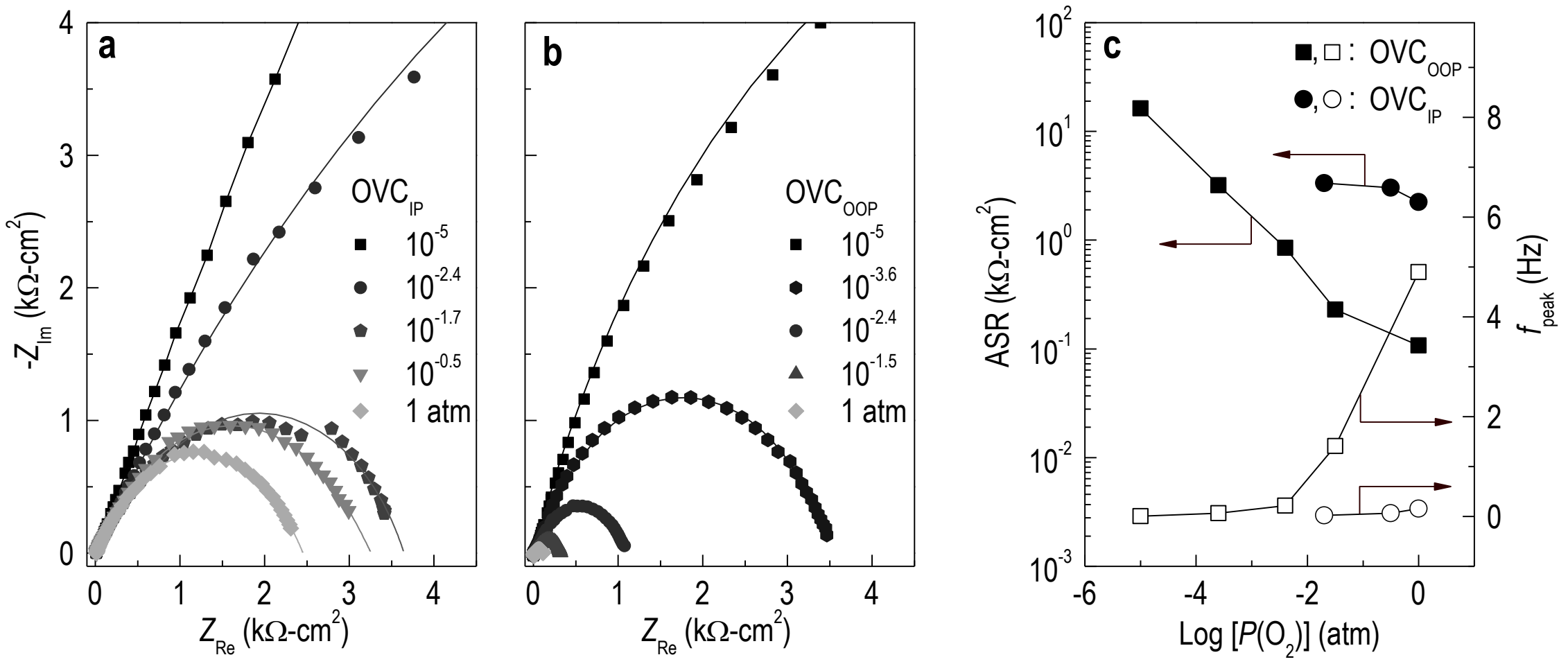

**Figure 2.** Oxygen partial pressure dependent impedance spectra. Nyquist plots of a) an $OVC_{IP}$ sample and b) an $OVC_{OOP}$ sample at 550 °C as a function of $P(O_2)$. c) Area specific resistance (ASR) and peak frequency ($f_{peak}$) are displayed as a function of $P(O_2)$. Due to extremely low catalytic activity in $OVC_{IP}$ BM-SCO, the ASR and $f_{peak}$ at low oxygen pressure were not determined.

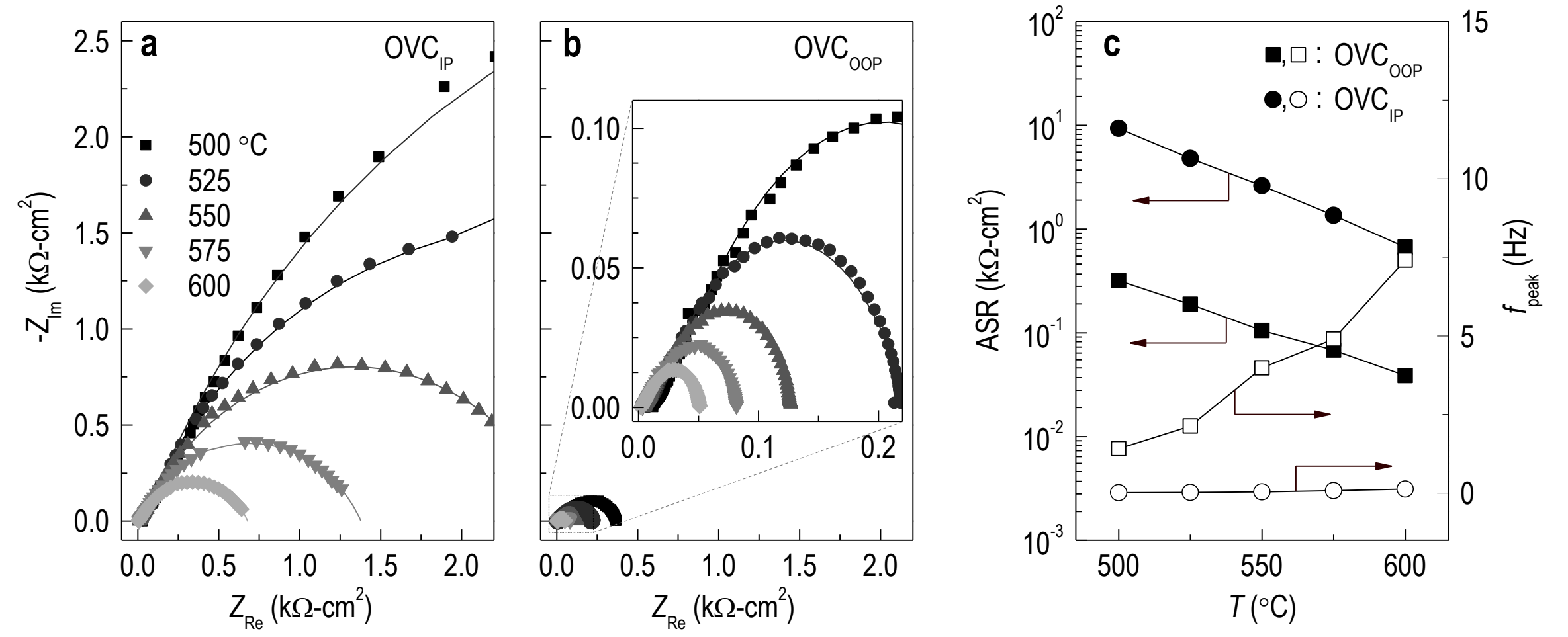

**Figure 3.** Temperature dependent impedance spectra. Nyquist plots of a) an $OVC_{IP}$ sample and b) an $OVC_{OOP}$ sample as a function of temperature ($500 \leq T \leq 600$ °C) measured in air. c) ASR and $f_{peak}$ as a function of temperature. The temperature dependent ASR data fitting (not shown) gives rise to thermal activation energies of $E_a$ ($OVC_{IP}$) = 1.5 eV and $E_a$ ($OVC_{OOP}$) = 1.2 eV.

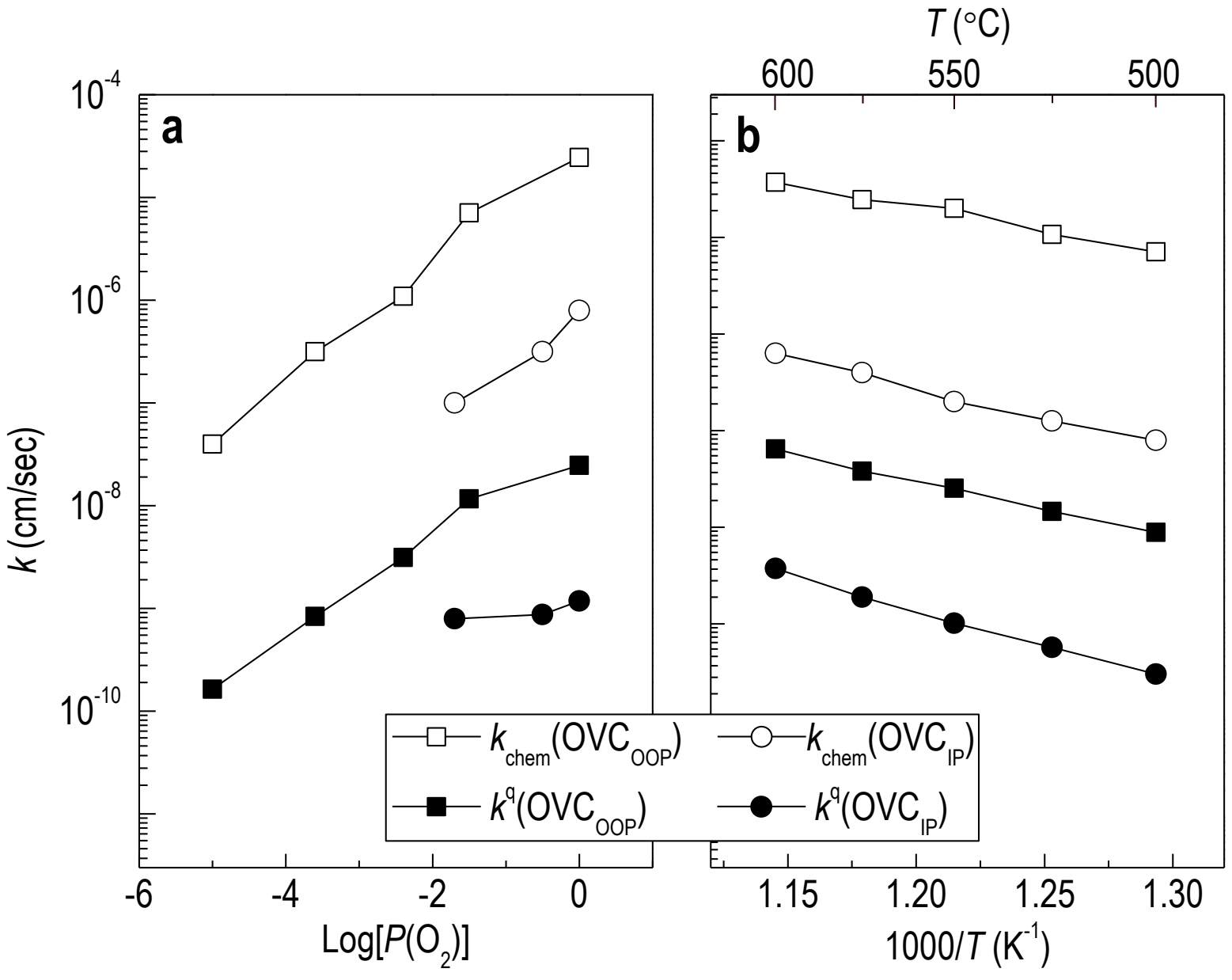

**Figure 4.** Surface exchange coefficients. a) $P(O_2)$ and b) temperature dependent electrical ($k^q$) and mean chemical ($k_{chem}$) surface exchange coefficients of the $OVC_{IP}$ and $OVC_{OOP}$ films. Highly enhanced ORR can be found in the $OVC_{OOP}$ sample.

# Supporting Information

## Epitaxy of $SrCoO_{2.5}$ thin films on YSZ substrates

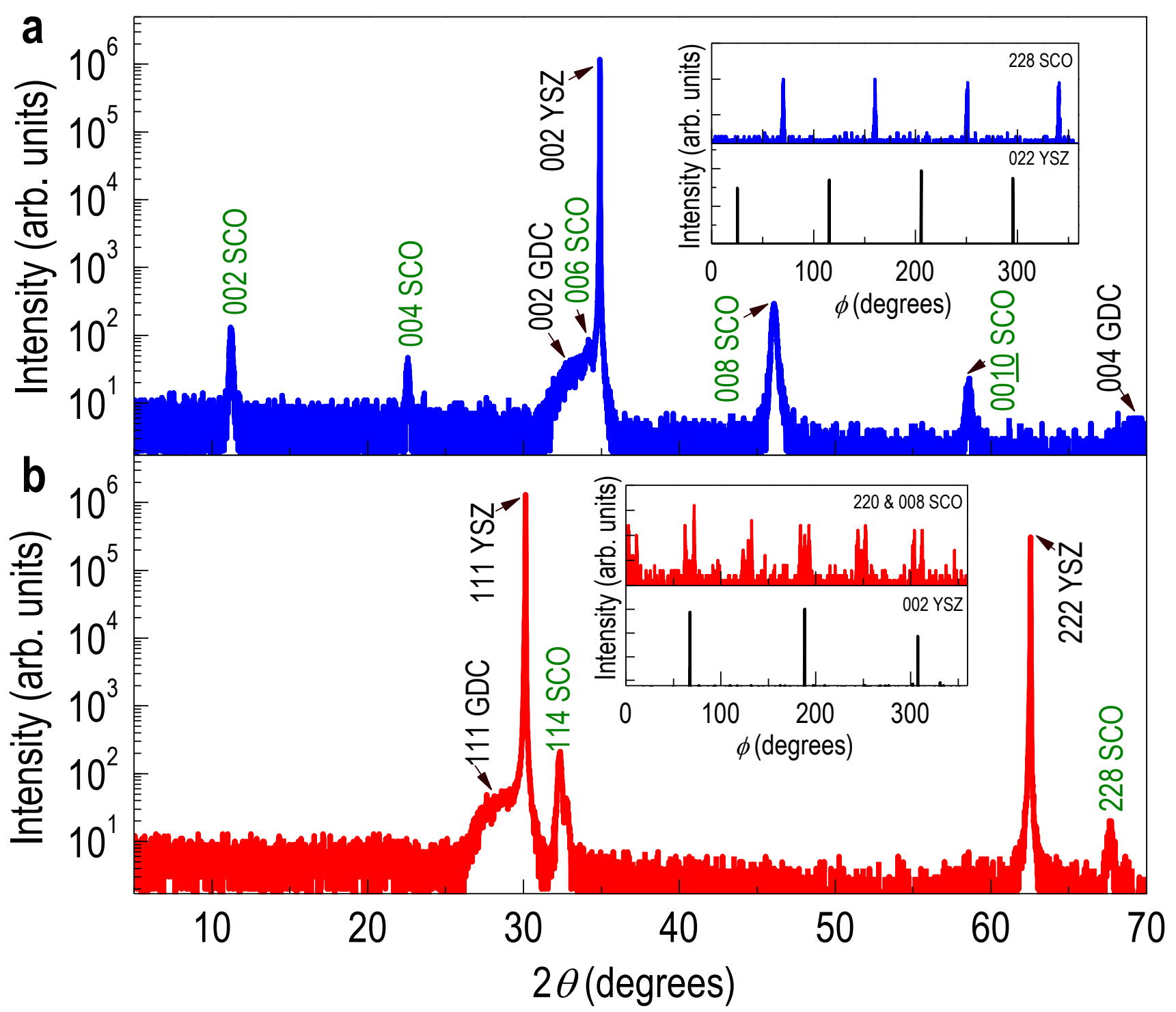

**Figure S1.** a) XRD $\theta$–2$\theta$ scan of a (001) BM-SCO film on a GDC-buffered (001) YSZ substrate. Inset: $\phi$ scans of the 228 BM-SCO and 022 YSZ reflections measured at $\psi$ = 45.0° and 45.0°, respectively, showing a cube-on-cube relationship. b) XRD $\theta$–2$\theta$ scan of a (114) BM-SCO film on a GDC-buffered (111) YSZ substrate. Inset: $\phi$ scans of the 220/008 BM-SCO and 002 YSZ reflections measured at $\psi$ = 45° and 54°, respectively.

(001)- and (114)-oriented brownmillerite $SrCoO_{2.5}$ epitaxial thin films were grown by pulsed laser epitaxy on Gd:$CeO_2$ (GDC)-buffered (001) and (111) YSZ substrates (see Figure S1 for XRD $\theta$-2$\theta$ scan results). Half order peaks from the (001) BM-SCO film on (001) YSZ from XRD $\theta$-2$\theta$ scans unambiguously confirm the successful stabilization of the vacancy ordered brownmillerite SCO phase. These orientations and vacancy ordering were further confirmed by performing XRD $\phi$ scans as shown in the inset of Figure S1: The $\phi$ scans shown in Figure S1a are for the 228 BM-SCO and 022 YSZ peaks recorded at tilted angles $\psi$ = 45.0° for both cases. For the (114) oriented BM-SCO film on YSZ (111), the $\phi$ scans in Figure S1b were recorded for the 220/008 BM-SCO and 002 YSZ peaks at $\psi \approx$ 45° and 54°, respectively. The off-axis XRD scans for (114) BM-SCO films clearly demonstrate the stabilization of the vacancy ordered brownmillerite phase. Therefore, the XRD results can be expressed as the following epitaxial relationships for the two samples (note GDC has the same orientation relation as YSZ):

$$(001)\,\text{BM-SCO} \parallel (001)\,\text{YSZ};\ [100]/[010]\,\text{BM-SCO} \parallel [100]\,\text{YSZ} \quad (1)$$

$$(114)\,\text{BM-SCO} \parallel (111)\,\text{YSZ};\ [\bar{4}01]\,\text{SCO} \parallel [\bar{1}01]\,\text{YSZ};\ [0\bar{4}1]\,\text{SCO} \parallel [0\bar{1}1]\,\text{YSZ} \quad (2)$$

From the epitaxial relationships, we could calculate the lattice mismatches of ~7% for the (001) BM-SCO film on YSZ (001) and ~6% for the (114) BM-SCO on YSZ (111). (Note that $SrCoO_{2.5}$ is orthorhombic with lattice constants of $a_o$ = 5.5739, $b_o$ = 5.4697, and $c_o$ = 15.7450 Å, while YSZ is cubic with $a_c$ = 5.140 Å.) Despite the relatively large lattice mismatches, high quality thin films were stabilized – FWHM values in rocking curve $\omega$ scans for the 008 and 114 BM-SCO peaks on (001) and (111) YSZ are 0.7° and 0.4°, respectively.

**Electrochemical impedance measurements and the specific rate-determining step**

Based on the superior catalytic property from the $OVC_{OOP}$ BM-SCO surface, we focused on the $OVC_{OOP}$ sample for determining the detailed specific rate-determining step in the ORR, which may give a hint to elucidate the unusually high catalytic activity on this $OVC_{OOP}$ BM-SCO surface. In general, the oxygen reduction reaction consists of several steps, which include the absorption of molecular oxygen, dissociation of molecular oxygen to atomic oxygen, and oxygen reduction involving charge transfer[3]. These steps can be identified by fitting the impedance data. The slope in a log-log plot of oxygen partial pressure dependent inversed resistance graph, i.e. the $n$ value in $1/R \propto P(O_2)^n$, expresses the rate determining step[25]. We checked this power law behavior using ASR of the $OVC_{OOP}$ BM-SCO film shown in Figure S2f. The $n$ value of the $OVC_{OOP}$ BM-SCO film is ~ 0.45, which indicates that rate determining step is dissociation of molecular oxygen to atomic oxygen[25]. It is worth noting that, in case of most doped-cobaltite bulks, the rate-determining step is known to be the charge transfer process for catalytic activity, while in case of doped-cobaltite thin films, it is the dissociative absorption[25, 26]. Therefore, our new observation implies that the ORR kinetics can be strongly influenced by the surface vacancy concentration controlled by the multivalent cobalt ions (i.e. octahedra and tetrahedral), which lead to a modification of the catalytic ORR process.

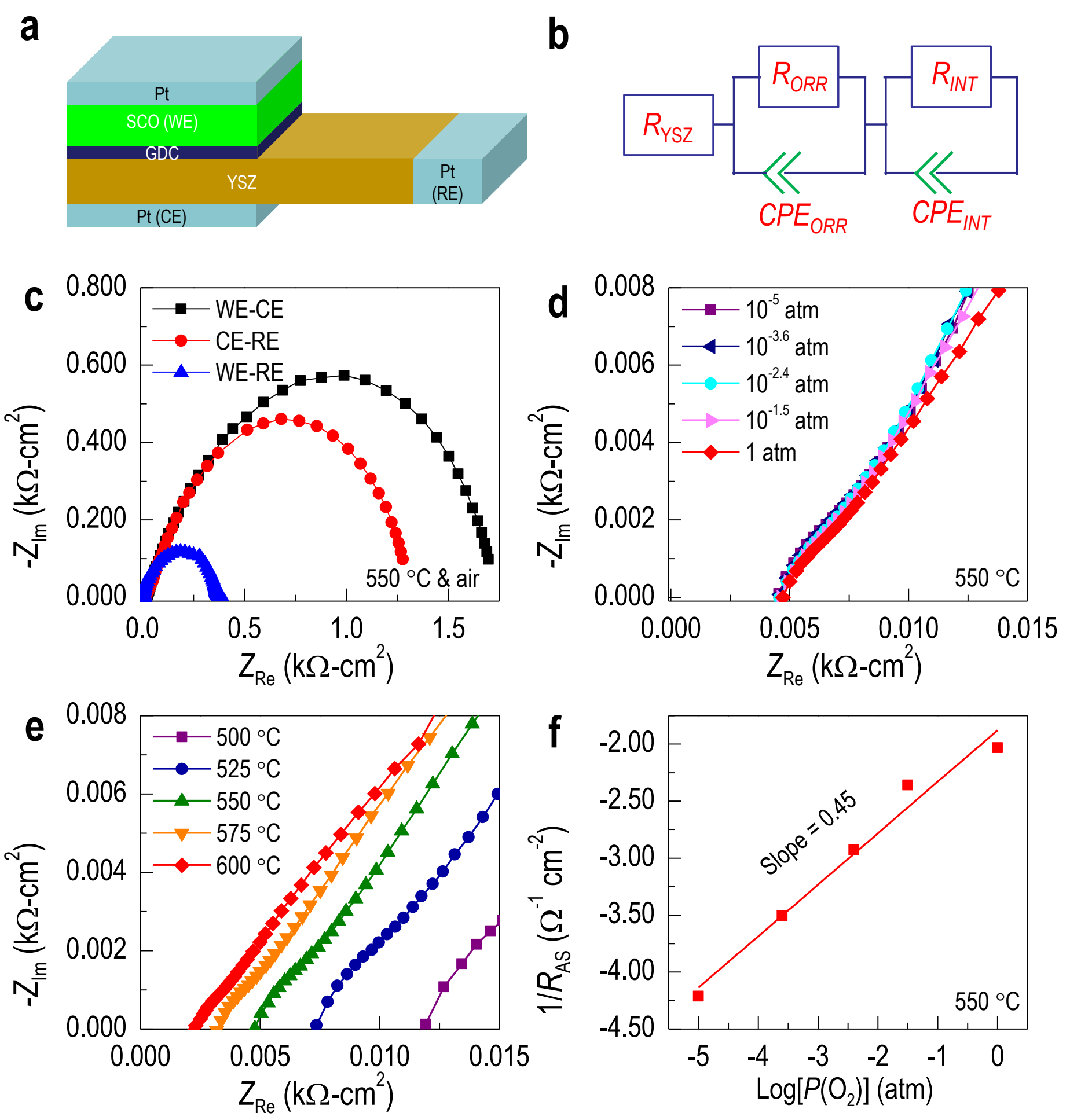

**Figure S2.** a) Schematic of the three-electrode method for impedance measurements and b) An equivalent circuit used to investigate the ORR kinetics ($R_{YSZ}$: YSZ electrode resistance at high frequencies, $R_{INT}$: interface resistance between the electrolyte and electrode, $R_{ORR}$: ORR resistance, and CPE: constant phase element). Note that WE, CE, and RE stand for work electrode, counter electrode, and reference electrode, respectively. c) Nyquist plots of impedance spectra measured for different electrode contacts with the $OVC_{OOP}$ sample. By comparing the data from the three different measurements, we could reliably exclude the contribution from the Pt counter electrode to the impedance spectra [i.e., $Z_{RE}$ (WE-RE) ≈ $Z_{RE}$ (WE-CE) – $Z_{RE}$ (CE-RE)]. Note that our impedance data shown in the main text are from measurements between WE-RE. d) $P(O_2)$ dependent Nyquist plots show unchanged non-zero real impedance values at high frequencies related with the oxygen ion conduction in YSZ. e) Temperature dependent Nyquist plots additionally support the high frequency non-zero values caused by the oxygen ion conduction in YSZ. f) Power law dependence of inverse of ASR from ORR as a function of $P(O_2)$.